\documentclass[a4paper,12pt]{article}
\usepackage{epsfig}
\usepackage{amsmath}
\usepackage{amssymb}
\usepackage{epsf}
\usepackage{graphicx}
\usepackage{cite}
\usepackage{graphics}
\usepackage{ulem}
\usepackage{rotating}

\newdimen\nude\newbox\chek
\def\slash#1{\setbox\chek=\hbox{$#1$}\nude=\wd\chek#1{\kern-\nude/}}

\newcommand{\sqrtsnn}{\sqrt{s_{_{{\rm N N}}}}}
\newcommand{\sqrts}{\sqrt{s}}
\newcommand{\produc}{{\rm prod}}
\newcommand{\jpsi}{J/\psi}
\newcommand{\sqrtspsin}{\sqrt{s_{_{\jpsi {\rm N}}}}}
\newcommand{\mpsi}{m_{_{\jpsi}}}
\newcommand{\mpsip}{m_{_{\psi'}}}
\newcommand{\mn}{m_{_{\rm N}}}
\newcommand{\A}{{\rm A}}
\newcommand{\B}{{\rm B}}
\newcommand{\TA}{T_{_\A}}
\newcommand{\dd}{{\rm d}\,}
\newcommand{\ndf}{{\rm ndf}}
\newcommand{\sig}{\sigma_{_{J/\psi {\rm N}}}}
\newcommand{\sigb}{\bar{\sigma}_{_{J/\psi {\rm N}}}}
\newcommand{\sighat}{\hat{\sigma}_{_{J/\psi {\rm N}}}}

\newcommand{\sigsh}{\sigma^{\rm nPDF}_{_{J/\psi {\rm N}}}}
\newcommand{\signDS}{\sigma^{{\small {\rm nDS}}}_{_{J/\psi {\rm N}}}}
\newcommand{\signDSg}{\sigma^{{\small {\rm nDSg}}}_{_{J/\psi {\rm N}}}}
\newcommand{\sigEKS}{\sigma^{{\small {\rm EKS}}}_{_{J/\psi {\rm N}}}}
\newcommand{\sigHKM}{\sigma^{{\small {\rm HKM}}}_{_{J/\psi {\rm N}}}}
\newcommand{\lqcd}{\Lambda_{_{_{\rm QCD}}}}

\def\pt{p_{_\perp}}
\def\X{{\rm X}}

\begin{document}

\setcounter{footnote}{3}
\renewcommand{\thefootnote}{\fnsymbol{footnote}} 	

\begin{flushright}
CERN-PH-TH/2006-189\\
LAPTH-1163/06\\
\texttt{hep-ph/0612043}\\

\end{flushright}

%%%%%%%%%%%%%%%%%%%%%%%%%%%%%%%%%%%%%%%%%%%%%%%%%%%%%%%%%%%%%%%%%%%%%%%%%%%%
\begin{center}
{\Large\bf A systematic study of $\jpsi$ suppression \\[0.3cm] in cold nuclear matter}
\end{center}

\begin{center}
{\large  Fran\c{c}ois Arleo$^{1}$\footnote{Email address: \texttt{arleo@cern.ch} (corresponding author)}\footnote{On leave from Laboratoire d'Annecy-le-Vieux de Physique Th\'eorique (LAPTH), UMR 5108 du CNRS associ\'ee \`a l'Universit\'e de Savoie, B.P. 110, 74941 Annecy-le-Vieux Cedex, France} and Vi-Nham Tram$^2$\footnote{Email address: \texttt{vntram@lbl.gov}}\footnote{Present address: Lawrence Berkeley National Laboratory, 1 Cyclotron Road, Berkeley CA, USA 94720}}\\[0.5cm]
$^1$ {\it CERN, PH Department, TH Division\\
1211 Geneva 23, Switzerland} \\~\\
$^2$ {\it Laboratoire Leprince-Ringuet, Ecole Polytechnique \\ 91128 Palaiseau Cedex, France}\\
\end{center}
%%%%%%%%%%%%%%%%%%%%%%%%%%%%%%%%%%%%%%%%%%%%%%%%%%%%%%%%%%%%%%%%%%%%%%%%%%%%

\begin{abstract}
Based on a Glauber model, a statistical analysis of all mid-rapidity $\jpsi$ hadroproduction and leptoproduction data on nuclear targets is carried out. This allows us to determine the $\jpsi$--nucleon inelastic cross section, whose knowledge is crucial to interpret the $\jpsi$ suppression observed in heavy-ion collisions, at SPS and at RHIC. The values of $\sig$ are extracted from each experiment. A clear tension between the different data sets is reported. The global fit of all data gives $\sig=3.4\pm 0.2$~mb, which is significantly smaller than previous estimates. A similar value, $\signDS=3.5\pm 0.2$~mb, is obtained when the nDS nuclear parton densities are included in the analysis, although we emphasize that the present uncertainties on gluon (anti)shadowing do not allow for a precise determination of $\sig$. Finally, no significant energy dependence of the $\jpsi$--N interaction is observed, unless strong nuclear modifications of the parton densities are assumed.
\end{abstract}

\setcounter{footnote}{0}
\renewcommand{\thefootnote}{\arabic{footnote}}

%%%%%%%%%%%%%%%%%%%%%%%%%%%%%%%%%%%%%%%%%%%%%%%%%%%%%%%%%%%%%%%%%%%%%%%%%%%%

\section{Introduction}

Twenty years ago, Matsui and Satz suggested the suppression of heavy-quark bound states in heavy-ion collisions as a sensitive probe for quark--gluon plasma (QGP) formation~\cite{Matsui:1986dk}. Since then, an intense experimental activity has been carried out. Soon after the Matsui and Satz proposal, the first measurements of $J/\psi$  production in heavy-ion reactions were performed in O--U and S--U collisions by the NA38 fixed-target experiment at the CERN Super Proton Synchrotron (SPS)~\cite{Baglin:1990ivBaglin:1991wi}. These were followed a few years later by the more precise NA50 data in Pb--Pb collisions~\cite{Abreu:1997jhAbreu:2000ni}, and more recently by the NA60 preliminary results in In--In collisions~\cite{Arnaldi:2006ee}, at a similar energy ($\sqrtsnn\simeq 20$~GeV). Finally, the PHENIX collaboration reported on data on $\jpsi$ production in Au--Au~\cite{Adler:2003rcxzAdare:2006ns} and on preliminary measurements in Cu--Cu scattering~\cite{Gunji:2006pc} at the Relativistic Heavy Ion Collider (RHIC) energy, $\sqrtsnn=200$~GeV. Remarkably, all these experimental results indicate a significant $J/\psi$ suppression in heavy ions (properly normalized by the number of NN binary collisions) with respect to $p$--$p$ scattering. 

However, it rapidly became clear that a key requirement to interpret these data is the quantitative understanding of all nuclear effects --~i.e. those that are not due to QGP formation~-- that could also affect $J/\psi$ production in nucleus--nucleus reactions. Among them, the nuclear modifications of the parton densities, measured in deep inelastic scattering (DIS) and Drell--Yan (DY) reactions, could affect the nuclear dependence of $\jpsi$ production. Another effect is the inelastic rescattering of the $J/\psi$ state --~or more generally a $c\bar{c}$ pair\footnote{Throughout this paper, we will abusively mention the ``$\jpsi$--N interaction''; yet at high energy what propagates through the nucleus is a $c\bar{c}$ pair. Similarly, we later denote by ``$\sig$'' the parameter governing the suppression of $\jpsi$ states in nuclear collisions, whether or not the $\jpsi$ is formed inside or outside the nuclear medium. The issue on the finite $\jpsi$ formation time is addressed in Sect.~\ref{sec:discussion}.}~-- in cold nuclear matter, which is expected to give the dominant contribution to the $J/\psi$ suppression in proton--nucleus and peripheral nucleus--nucleus collisions. The strength of this mechanism is monitored by essentially one physical parameter, the $J/\psi$--nucleon inelastic cross section, $\sig$. 

On the theoretical side, this cross section has been computed perturbatively in QCD in the leading-twist approximation~\cite{Peskin:1979vaBhanot:1979vb} and proves of the order of a few millibarns in the $J/\psi$--N energy range of interest~\cite{Kharzeev:1994pz,Arleo:2001mp} (similar estimates based on a generalized vector dominance model are given in~\cite{Hufner:1997jg}). However, the relative smallness of the charm quark mass (with respect to $\lqcd$) may lead to significant power corrections; consequently these numbers should not be taken at face value but rather seen as {\it qualitative} estimates only. It is therefore necessary to rely on existing experimental data to determine this quantity. Fortunately, the measurement of $J/\psi$ production on nuclear targets has been performed extensively for hadroproduction (E537~\cite{Katsanevas:1987pt}, NA3~\cite{Badier:1983dg}, NA38~\cite{Abreu:1998ee}, NA50~\cite{Alessandro:2003pc}, NA60~\cite{Scomparin:2006pc}, E672~\cite{Kartik:1990it}, E772~\cite{Alde:1990wa}, E866~\cite{Leitch:1999ea}, HERA-B~\cite{Husemann:2005yq}, PHENIX~\cite{Adler:2005ph}), photoproduction (SLAC~\cite{Anderson:1976hi}, E691~\cite{Sokoloff:1986bu}) and leptoproduction (EMC~\cite{Aubert:1984br}, NMC~\cite{Amaudruz:1991sr}). In the spirit of the Gerschel and H\"ufner analysis~\cite{Gerschel:1993uh} and the subsequent work by Kharzeev, Louren\c{c}o, Nardi and Satz~\cite{Kharzeev:1996yx}, it is the aim of this paper to extract the inelastic $\jpsi$--N cross section systematically from a Glauber analysis of these experimental results --~including the possible nuclear corrections to the parton distribution functions~-- and thus to provide a baseline of the $\jpsi$ absorption mechanism in nuclear matter from all available data.

The outline of the paper is the following. The $\jpsi$ production cross section in nuclei is detailed in Section~\ref{sec:production}. The method adopted in this analysis is then described in Section~\ref{sec:method}, while Section~\ref{sec:data} is devoted to the data selection. The results obtained are presented in Section~\ref{sec:results}. Section~\ref{sec:shadowing} is devoted to a systematic investigation of the uncertainties coming from the nuclear parton densities, before the results are discussed in Sect.~\ref{sec:discussion}. Finally, a summary of what has been carried out here is given (Section~\ref{sec:summary}).

%%%%%%%%%%%%%%%%%%%%%%%%%%%%%%%%%%%%%%%%%%%%%%%%%%%%%%%%%%%%%%%%%%%%%%%%%%%%

\section{$\jpsi$ production in nuclear targets}
\label{sec:production}

\subsection{Production}

The charmonium production cross section in hadron ($h$) -- nucleus (A) collisions is determined in this work within the Colour Evaporation Model (CEM). Neglecting the possible inelastic interaction of the produced $J/\psi$ in nuclear matter, the leading order (LO) production cross section, as a function of the longitudinal momentum fraction, $x_{_{\rm F}}$, reads~\cite{Barger:1979jsBarger:1980mg} 
\begin{eqnarray}
  \label{eq:cem}
  \frac{\dd \sigma^{\produc}}{\dd x_{_{\rm F}}}(h \ \A \to J/\psi \ \X) &=& \rho_{_{\jpsi}} \int_{2 m_{_c}}^{2 m_{_D}} \dd{m} \ \frac{2 m}{\sqrt{x_{_{\rm F}}^2 s + 4 m^2}} \ \Bigg[ f_g^{h}(x_{_1},m^2) \ f_g^{\A}(x_{_2},m^2) \  \sigma_{gg}(m^2) \nonumber \\[0.6cm]
&+& \sum_{q=u,d,s} \bigg\{ f_q^{h}(x_{_1},m^2) \ f_{\bar{q}}^{\A}(x_{_2},m^2)  + f_{\bar{q}}^{h}(x_{_1},m^2) \ f_q^{\A}(x_{_2},m^2) \bigg\} \sigma_{q\bar{q}}(m^2) \Bigg],\nonumber \\ & &
\end{eqnarray}
where 
\begin{equation*}
  x_{_{1, 2}} = \frac{1}{2} \, \left( \sqrt{x_{_{\rm F}}^2+4 \ m^2/s} \pm x_{_{\rm F}} \right)
\end{equation*}
are the projectile and target-parton momentum-fractions ($\sqrt{s}$ being the centre-of-mass energy of the hadronic collision), $m_{_c}=1.2$~GeV (respectively, $m_{_D}=1.87$~GeV) is the charm-quark ($D$-meson) mass, and $\sigma_{q\bar{q}}$ (respectively, $\sigma_{gg}$) is the LO $c\bar{c}$ partonic production cross section in the quark--antiquark annihilation $q\bar{q}$ (gluon fusion, $gg$) channel. In this model, the non-perturbative transition from the $c\bar{c}$ pair to the $\jpsi$ state is hidden in the proportionality factor, $\rho_{_{\jpsi}}\simeq 0.5$, which is adjusted to the experimental measurements~\cite{Amundson:1996qr}. Note that, since only cross-section {\it ratios} of heavy to light nuclei will be considered in the following, this factor proves irrelevant in the present analysis. In Eq.~(\ref{eq:cem}), $f_{i}^{h}$ and $f_{i}^{\A}$ stand respectively for the parton distribution function (PDF) in the hadron and in the nucleus. The PDF in a nucleus with $Z$ protons and the atomic mass number $A$ is written as the sum of the proton ($f_{i}^{p/\A}$) and the neutron ($f_{i}^{n/\A}$) parton densities in a nucleus:
\begin{equation}
  \label{eq:isospin}
  f_{i}^{\A} = Z \ f_{i}^{p/\A} + (A-Z) \ f_{i}^{n/\A},
\end{equation}
where $f_{i}^{n/\A}$ is obtained from $f_{i}^{p/\A}$ by isospin conjugation: $u^{n/A}=d^{p/\A}$, $d^{n/\A}=u^{p/\A}$, $s^{n/\A}=s^{p/\A}$. This isospin effect only affects the $J/\psi$ production cross section in nuclei marginally, as long as gluon fusion is the dominant channel. However, whenever the projectile $h$ contains valence antiquarks (e.g. $\bar{p}$, $\pi^\pm$), the $\bar{q}^h_{_V}q^{\A}_{_V}$ annihilation process becomes dominant, provided $\sqrt{s}$ is not too large with respect to $m_{_{\jpsi}}$ (say, $\sqrt{s}\lesssim~20$~GeV), and significant effects are expected because of the very different valence-quark distribution in the proton and in the neutron.

In virtual photon reactions, the $\jpsi$ inelastic production cross section is computed in QCD to leading order according to Refs.~\cite{Berger:1980niMartin:1987ww}:
\begin{equation}
  \label{eq:eprod}
   \frac{\dd \sigma^{\produc}}{\dd x} (\gamma^* \ \A \to J/\psi \ \X)  =  \frac{\Gamma_{ee}}{\mpsi^3} \ f(x, s/\mpsi^2) \ x f_g^{\A}(x, Q^2),
\end{equation}
with $s$ being the square of the $\gamma^*$--N centre-of-mass energy, and where $\Gamma_{ee}=4.7$~keV is the $\jpsi$ leptonic width and $f(x, s/\mpsi^2)$ is a sharply peaked function above $\mpsi^2/s$~\cite{Berger:1980niMartin:1987ww}. Just as $\rho_{_{\jpsi}}$ in (\ref{eq:cem}), these two factors cancel out in the heavy-to-light nucleus cross-section ratio. We shall therefore only exploit the fact that the $J/\psi$ inelastic leptoproduction in nuclei is proportional to the gluon density in the nuclear target, $f_g^{\A}(x, Q^2)$, where the hard scale $Q$ is given by the $J/\psi$ mass.

The PDFs in Eqs.~(\ref{eq:cem}) and (\ref{eq:eprod}) are taken from the LO parametrizations CTEQ6L for the proton~\cite{Pumplin:2002vw} and SMRS for the pion~\cite{Sutton:1991ay}. The nuclear PDFs are discussed in the next section.

\subsection{Nuclear parton densities}
\label{sec:npdf}

The nuclear parton densities are known to differ somehow from those in a proton, $f_i^{p/\A}(x) \neq f_i^p(x)$ (see~e.g.~\cite{Armesto:2006ph}, and references therein, for a recent review) on the whole Bjorken-$x$ range. At small $x\lesssim 0.05$, nuclear PDFs (nPDF) are depleted, $f_i^{p/\A}(x) < f_i^p(x)$, in the so-called shadowing region,  while they may be slightly enhanced (``antishadowing'') in an intermediate $x\simeq 0.05$--$0.2$ range. Above $x\gtrsim 0.2$, the quark distributions were determined from the EMC muoproduction data on nuclei~\cite{Aubert:1983xm} and proved smaller than the quark densities in a proton (depletion usually referred to as the ``EMC effect''). At very large $x\simeq 1$, $f_i^{p/\A}(x) \gg f_i^p(x)$ because of the Fermi motion of the nucleons in the nuclei.

\begin{figure}[htb]
  \begin{center}
    \includegraphics[height=9.0cm]{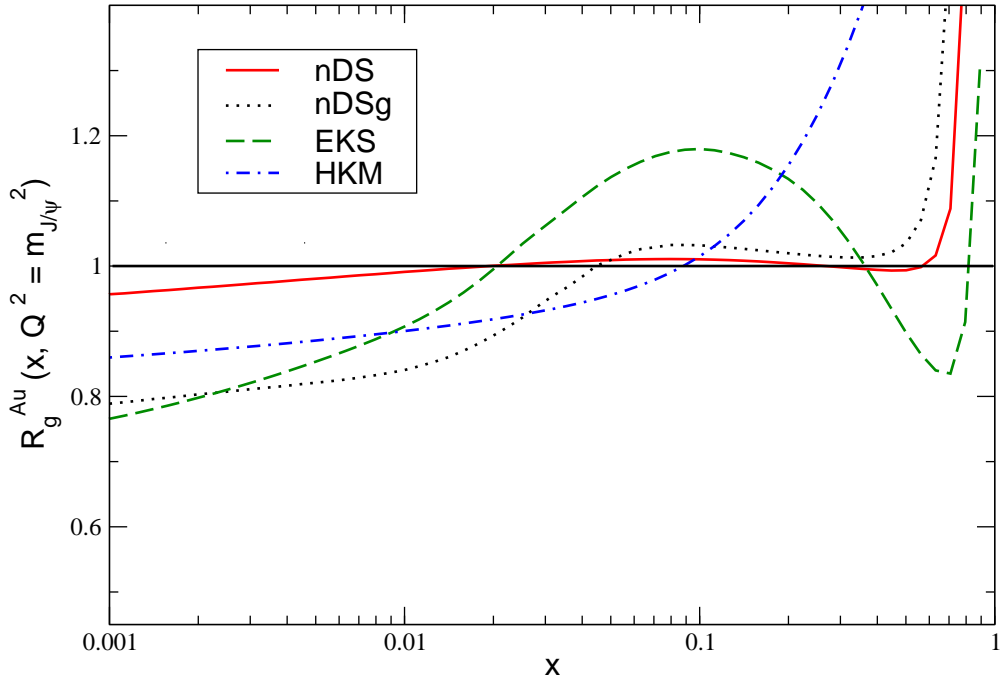}
  \end{center}
\caption{The ratio of the gluon distribution in a gold nucleus over that in a proton, $R_g^{\rm Au}(x, \mpsi^2$), plotted as a function of Bjorken-$x$ using the nDS, nDSg, EKS and HKM parametrizations.}
  \label{fig:npdf}
\end{figure}

Using DIS and DY measurements on various nuclear targets, several LO fits of the nuclear parton densities, analogous to the standard QCD fits in a proton, have been performed, first by Eskola et al. (EKS)~\cite{Eskola:1998iyEskola:1998df}, and more recently by Hirai et al. (HKM)~\cite{Hirai:2001npHirai:2004wq} and De Florian and Sassot (nDS)~\cite{deFlorian:2003qf}. Note that in the latter analysis, a parametrization at NLO accuracy is also given. Unfortunately, the current precision (and variety) of the nuclear data does not yet allow the nPDFs to be as constrained as those in the proton, leading in particular to rather large uncertainties (say, $20\%$ in the $x$ range of interest here) in the gluon sector. This is illustrated in Fig.~\ref{fig:npdf}, where the ratio of the gold nucleus over the proton PDF at the $\jpsi$ mass scale, $R_g^{\rm Au}(x, Q^2=\mpsi^2)$, is plotted as a function of $x$ using the different parametrizations.

Since the nuclear dependence of the gluon and the quark densities are a priori not identical, $R_g^{\rm A}(x, \mpsi^2)\ne R_q^{\rm A}(x, \mpsi^2)$, it is necessary to determine the relative proportion of the $gg$ fusion and $q\bar{q}$ channel in the $\jpsi$ hadroproduction process, which is given here by the CEM, Eq.~(\ref{eq:cem}). However, we believe that this proportion --~and hence our predictions~-- should not depend on the specific model assumed for the heavy-quarkonium production mechanism (Colour Singlet Model, Colour Evaporation Model, Non-Relativistic QCD, \dots); see e.g.~\cite{Vogt:2004dhVogt:2005ia}.

The parton distributions in nuclei not being precisely fixed yet by the current experiments, the analysis is first carried out without any nuclear corrections to the PDFs, assuming $f_i^{p/\A}(x) = f_i^p(x)$. The analysis is also performed using the LO nDS parametrization, whose agreement with DIS and DY nuclear data is the best among the different nPDF sets~\cite{deFlorian:2003qf}. Furthermore, we shall critically discuss in Section~\ref{sec:shadowing} how the uncertainties on gluon (anti)shadowing affect our results. 

\subsection{Nuclear absorption}
\label{sec:nucabs}

The factorization between the charmonium production process and the subsequent possible $J/\psi$ inelastic interaction with nuclear matter is assumed. Quite generally, we may thus write the $J/\psi$ production cross section as
\begin{equation}
  \label{eq:xs}
  \frac{\dd \sigma }{\dd x}\left(h,\gamma^* \ \A \to J/\psi \ \X\right) = S_{\rm{abs}}(\A\ , \sig) \times \frac{\dd \sigma^{\produc} }{\dd x}\left(h,\gamma^* \ \A \to J/\psi \ \X\right),
\end{equation}
where $S_{\rm{abs}}(\A\ , \sig)$ denotes the probability for no interaction (or ``survival probability'') of the $J/\psi$ meson with the target nucleus. It depends of course on both the atomic mass number $A$ of the nucleus and the $J/\psi$--N inelastic cross section, $\sig$, and is given in a Glauber model by\cite{Capella:1988ha}
\begin{equation}
  \label{eq:supp}
  S_{\rm{abs}}(\A, \sig) = \frac{1}{(A-1) \ \sig}\, \int \dd {\bf b} \left( 1 - e^{- (1-1/A) \ \TA({\bf b}) \ \sig} \right),
\end{equation}
with the thickness function $\TA({\bf b})$ 
\begin{equation}
  \label{eq:tf}
  \TA({\bf b}) = \int_{-\infty}^{+\infty} \dd z \ \rho({\bf b}, z).
\end{equation}
The thickness function Eq.~(\ref{eq:tf}) is determined for all nuclei using a 2-parameter Fermi model for the nuclear density profile $\rho({\bf b}, z)$, except for the Be, C, and Ca targets where a harmonic oscillator, a sum of gaussians, and a 3-parameter Fermi model is used respectively (see Ref.~\cite{DeJager:1987qc} for details) and for the Ag and Pt nuclei whose density profiles, missing in the analysis~\cite{DeJager:1987qc}, are taken from the parametrization given by the FRITIOF Monte Carlo code~\cite{Pi:1992ug}. Finally, we neglect the nuclear absorption in the proton and deuterium targets, that is $S_{\rm{abs}}(A\le 2\ , \sig)=1$.

%%%%%%%%%%%%%%%%%%%%%%%%%%%%%%%%%%%%%%%%%%%%%%%%%%%%%%%%%%%%%%%%%%%%%%%%%%%%

\section{Method}
\label{sec:method}

After discussing the theoretical calculations of $J/\psi$ hadroproduction and leptoproduction on nuclei, we present in this section the method followed in this study. The $J/\psi$--N inelastic cross section is extracted, for each experimental sample $\ell$ with $n_{_\ell}$ data points, from the minimization of the $\chi^2_{_\ell}$ function~\cite{Stump:2001gu}
\begin{equation}
  \label{eq:chi2}
  \chi^2_{_\ell}(\sig) = \sum_{i = 1}^{n_{_\ell}} \ \left[\frac{R_i^{\rm exp} - R_i^{\rm th}(\sig)}{\sigma_{i}}\right]^2 - {\bf V M^{-1} V},
\end{equation}
where the theoretical nuclear production ratio,
\begin{equation}
  \label{eq:ratio}
  R_i^{\rm th}(\sig) = \frac{B}{A_{_i}} \ \ \frac{\dd\sigma}{\dd x}\Bigg|_{x_{_{F_i}},\ x_{_{_i}}} \hspace{-0.45cm}\left( h,\gamma^* \ \A_{_i} \to \jpsi \ \X \right) \ \Biggr/ \ \frac{\dd\sigma}{\dd x}\Bigg|_{x_{_{F_i}},\ x_{_{_i}}} \hspace{-0.45cm}\left(h,\gamma^* \ \B \to \jpsi \ \X\right)
\end{equation}
computed from Eq.~(\ref{eq:xs}) at a given $x_{_{F_i}}$ or Bjorken $x_{_{_i}}$, depends explicitly on the free, but positive, parameter, $\sig$ (which we denote generically by $\sigsh$ when nuclear parton distributions are used in the theoretical calculations). In Eq.~(\ref{eq:chi2}), $\sigma_{i}$ represents the statistical and the uncorrelated systematic errors, added in quadrature, of the experimental nuclear production ratio, $R_{i}^{\rm exp}$, in a heavy nucleus ($\A_{_i}$) with respect to the lightest target available in the data sample\footnote{Note that for the NA3 experiment, the inverted ratio $R_{i}^{\rm exp}(p/{\rm Pt})$ is measured.} (B). In order to avoid too large systematic errors in the experimental ratio, both reactions on targets $\A_{_i}$ and B are required to be taken from the same experiment and at the same centre-of-mass energy. 

The {\it correlated} systematic error from source $k$ on the data point $i$, $\beta_{ik}$, appears in (\ref{eq:chi2}) through the $K$-dimensional vector
\begin{equation}
V_k \ =\ \sum_{i=1}^{n} \ \frac{\beta_{ik} \ [R_i^{\rm exp} - R_i^{\rm th}(\sig)]}{\sigma^2_i},
\end{equation}
and the $K\times K$ matrix 
\begin{equation}
M_{kl} \ =\ \delta_{kl} \ + \ \sum_{i=1}^{n} \ \frac{\beta_{ik} \ \beta_{il}}{\sigma^2_i},
\end{equation}
where $K$ is the number of distinct correlated errors. In the present work, we shall mostly be concerned by normalization errors (e.g. due to the uncertainty on the lightest nucleus production cross section), and the number of distinct errors in {\it each} experiment is $K=1$. In practice, ${\bf M}$ is a number when $\sig$ is extracted from one experiment, and a diagonal ${\cal N}\times{\cal N}$ matrix when performing a global fit on the ${\cal N}$ data samples.

The $\chi^2$ definition, Eq.~(\ref{eq:chi2}), is shown to be actually equivalent~(see \cite{Stump:2001gu}) to the more traditional approach, which involves the inversion of the $n\times n$ covariant matrix:
\begin{equation}
  \label{eq:cov}
  M^{\rm cov}_{ij} \ =\ \delta_{ij}\ \sigma_{i}^2\ + \ \sum_{k=1}^{K} \ \beta_{ik} \ \beta_{jk}.
\end{equation}
Since here $K \ll n$, we will use the former definition (\ref{eq:chi2}) in the present paper, which proves of course more efficient than Eq.~(\ref{eq:cov}). The phenomenological consequences of taking the {\it correlated} errors properly into account in this $\chi^2$ analysis --~instead of adding all errors quadratically~-- will be discussed in Sect.~\ref{sec:results}.

As already stressed, the estimated $\jpsi$--N cross section, $\sighat$, is determined from the minimization of the $\chi^2$ function:
\begin{equation}
  \label{eq:min}
  \chi^2(\sighat) = \min\left[\chi^2(\sig)\right] \equiv \chi_{_{\rm min}}^2.
\end{equation}
Including the correlated errors in the $\chi^2$ definition, the $1\sigma$ error $\delta\sighat$ on the fitted parameter\footnote{We drop the subscript ``min'' in the following, and whenever a $\chi^2$ value is quoted, it is understood to be $\chi_{_{\rm min}}^2$. The ``hat'' on the estimated cross section $\sighat$, and its error $\delta\sighat$, will also be removed in the following, for clarity.} $\sighat$ leads to a deviation of $\chi^2$ by one unit from its minimum: 
\begin{equation}
  \label{eq:dchi2}
  \Delta \chi^2 \equiv \chi^2(\sighat \pm \delta \sighat) - \chi_{_{\rm min}}^2 = 1.
\end{equation}
Note, however, that this criterion is no longer valid when correlations are neglected; in which case, it should be emphasized that $\Delta \chi^2$ can actually be much larger than 1~\cite{Stump:2001gu}. 

For some data samples, the agreement between the theoretical predictions and the experimental measurements is poor, leading to rather large $\chi^2\gtrsim~1$ values. Following the prescription from the Particle Data Group~\cite{Yao:2006px}, the $1\sigma$ error is rescaled,
\begin{equation}
\delta \sigb \ = \ {\rm S} \ \times \ \delta \sig,
\end{equation}
where the factor S is defined as:
\begin{equation}\label{eq:Sfactor}
{\rm S} \ \equiv \ \left[\chi^2\bigg/(n-1)\right]^{1/2} \quad {\mathrm{if}} \quad \chi^2/\ndf > 1, 
\end{equation}
and S~$\equiv 1$ otherwise. 

The $\jpsi$--N cross section is systematically determined in Sect.~\ref{sec:results} from the individual data samples. It will also be extracted from a global fit of all available data, from the minimization of the weighted $\chi^2$ function:
\begin{equation}
  \chi^2(\sig) = \sum_{\ell=1}^{\cal N} \ \ {\rm S}_{_\ell}^{-1} \ \ \chi_{_\ell}^2(\sig).
\end{equation}
This global fit analysis will thus give lesser importance to the data sets for which no good agreement between data and theory is observed.

%%%%%%%%%%%%%%%%%%%%%%%%%%%%%%%%%%%%%%%%%%%%%%%%%%%%%%%%%%%%%%%%%%%%%%%%%%%%

\section{Data selection}
\label{sec:data}

The data sets used in this analysis are summarized in Table~\ref{tab:datasets}, in which the centre-of-mass energy $\sqrts$ of the hadron--nucleon or photon--nucleon system, the $x_{_{\rm F}}$ domain\footnote{Bjorken $x$ is understood for the leptoproduction experiment.}, the projectile/target species and the number $n$ of points in each data sample are specified. 

\begin{table}[htb]
  \begin{center}
  \begin{tabular}[c]{p{1.8cm}cccccc}
    \hline
    \hline
    Exp. & Ref. & $\sqrt{s}$ (GeV) & $x_{_{\rm F}}$ & Proj. & Target &  No. data \\
    \hline
 & & & & & & \\
    E537 & \cite{Katsanevas:1987pt} & 15.3 &  0--0.25 & $\pi^-$, $\bar{p}$ & Be, Cu, W & 9 \\
    NA3 & \cite{Badier:1983dg} & 16.7--22.9 & 0--0.3 & $\pi^\pm$, $p$ & $p$, Pt & 21 \\
    NA38 & \cite{Abreu:1998ee} & 29.1 & 0.1 & $p$ & C, Al, Cu, W & 3 \\
    NA50 & \cite{Alessandro:2003pc} & 27.4--29.1 & $-0.075$--0.075 & $p$ & Be, Al, Cu, Ag, W, Pb & 13 \\
    E672 & \cite{Kartik:1990it} & 31.5 & 0.3 & $\pi^-$ & C, Al, Cu, Pb & 3 \\
    E866 & \cite{Leitch:1999ea} & 38.1 & $-0.1$--$0.3$ & $p$ & Be, Fe, W & 10 \\
    HERA-B & \cite{Husemann:2005yq} & 41.6 & $-0.3$--0.1 & $p$ & C, W & 6 \\
    PHENIX & \cite{Adler:2005ph} & 200 & $-0.1$--$0.1$ & $p$, d & $p$, Au & 5 \\
    NMC & \cite{Amaudruz:1991sr} & 8--21 & 0.05--0.15  & $\gamma^*$ & C, Sn & 6 \\
 & & & & & & \\
    \hline
    \hline
\end{tabular}
  \caption{Data sets selected for the present analysis.}
  \label{tab:datasets}
  \end{center}
\end{table}

\subsection{Hadroproduction data}
\label{sec:hadro}

Concerning hadroproduction measurements, we analyse the data from the E537, NA3, NA38, NA50, E672, E866, HERA-B, and PHENIX collaborations. The projectiles used were mainly protons (NA3, NA38, NA50, E866, HERA-B, PHENIX), but also pions (E537, NA3, E672), antiprotons (E537), and deuterium nuclei (PHENIX). The centre-of-mass energy range is $\sqrts=15$--$200$~GeV which, in the $x_{_F}$ domain covered by each experiment and which fits the kinematic requirements detailed below, translates into an $x_{_2}= 0.015$--$0.3$ window for $97\%$ of the data points\footnote{i.e. excluding the two forward measurements by PHENIX at $x_{_2}=2\ 10^{-3}$ and $x_{_2}=3.6\ 10^{-3}$.}. Note that the energy in the $\jpsi$-nucleon system, given by
\begin{equation}
  \label{eq:sqrts_psin}
  \sqrtspsin \simeq \mpsi \ \sqrt{\frac{1+x_{_2}}{x_{_2}}},
\end{equation}
has a rather narrow coverage, $\sqrtspsin \simeq  6.5{\rm -}25$~GeV, despite the wide range in $\sqrts$ probed by the various experiments. 

The E772 data~\cite{Alde:1990wa} were not taken into account in this analysis: the incomplete coverage of the E772 spectrometer apparatus induces, at small $x_{_{\rm F}}$, a strong correlation between the longitudinal (say, $x_{_{\rm F}}$) and the transverse ($\pt$) momentum of the produced $\jpsi$'s. Consequently, $\jpsi$ production occurs in this $x_{_{\rm F}}$ domain mainly at very small $\pt$, for which the nuclear dependence is the strongest because of the Cronin effect~(\cite{Leitch:2006pc} and M.~Leitch, private communication). This explains why the uncorrected $\jpsi$ suppression reported by E772 in~\cite{Alde:1990wa} is systematically more pronounced than the more precise and more recent E866 data, in the same kinematic domain, which are corrected for this bias~\cite{Leitch:1999ea}.

We would like to mention that the NA60 experiment has recently reported on a preliminary measurement of $\jpsi$ production in $p$--A collisions at $\sqrtsnn=17.3$~GeV~\cite{Scomparin:2006pc}, i.e. the very energy of the Pb--Pb collisions in which an anomalous $\jpsi$ suppression has been measured by NA50 (see Introduction). Unfortunately, no $\jpsi$ production ratio $R^{\rm exp}$ is available yet, which does not allow the present NA60 data point to be used in the present analysis. Obviously, the coming measurements in Be, Al, Cu, In, W, Pb, and U targets will shed important light on the nuclear dependence of $\jpsi$ suppression, and should be used in a future similar analysis.

Let us finally comment on the kinematic requirements imposed on the available hadroproduction data sets. The strong $J/\psi$ suppression reported at large $x_{_{\rm F}}$ in various hadroproduction experiments~\cite{Badier:1983dg,Alde:1990wa,Leitch:1999ea} is so far poorly understood. It may come from an interplay of several mechanisms such as inelastic interaction with nuclear matter, shadowing effects, and the onset of a new QCD production mechanism coming from the intrinsic charm-quark content of the projectile hadrons (see~\cite{Vogt:1999dw}, for instance, for a quantitative analysis of these various nuclear effects). In any case, it is unlikely that inelastic interaction proves the dominant process responsible for such a strong $J/\psi$ suppression. Furthermore, since the main motivation of our study is to have a handle on the expected amount of $\jpsi$ nuclear absorption in heavy-ion collisions, in which $\jpsi$'s are mainly produced at mid-rapidity, the large-$x_{_{\rm F}}$ region is discarded. Consequently, an arbitrary cut $x_{_{\rm F}} \le 0.3$ is therefore imposed on all hadroproduction data sets.

\subsection{Leptoproduction data}
\label{sec:lepto}

As for leptoproduction experiments, the NMC data~\cite{Amaudruz:1991sr} are analysed in this study. Using a 200~GeV and 280~GeV incident muon beam, the virtual photon energy $\nu$ ranges from 40 to 240~GeV in the laboratory frame, corresponding roughly to $\gamma^*$--N centre-of-mass energies $\sqrts=8$--21~GeV. The typical Bjorken-$x$ range probed in the gluon distributions of the nuclear targets is $x=0.05$--$0.15$ in the NMC kinematics. In order to remove the contribution of the $\jpsi$ coherent production, in which the nucleus as a whole interacts with the virtual photon, a minimal transverse momentum cut, $\pt^2\ge 0.4$~GeV$^2$, was applied. In addition, a cut on the energy fraction carried by the $J/\psi$ in the laboratory frame, $z=E_{_{J/\psi}}/\nu \le 0.85$, was applied by NMC to safely exclude quasi-elastic events~\cite{Amaudruz:1991sr}. On the contrary, the measurements from the EMC collaboration~\cite{Aubert:1984br} are not taken into account in this work, since their resolution in $z$ and $\pt$ did not allow them to subtract the coherent contribution and thus extract the genuine $J/\psi$ inelastic production,~Eq.~(\ref{eq:eprod}).

\subsection{Photoproduction data}
\label{sec:photo}

Regarding the photoproduction experiments, only quasi-elastic events were measured at SLAC~\cite{Anderson:1976hi}, which prevents us from using these data since the focus is put on $\jpsi$ inelastic production. Similarly, the lack of a $z$ energy-fraction cut in the E691 experiment at Fermilab does not allow for the extraction of the nuclear dependence of the inelastic events. Therefore, these measurements are not selected in the present analysis. Understanding quantitatively the nuclear dependence of $\jpsi$ quasi-elastic photoproduction (and leptoproduction) would certainly be interesting in its own, and will discussed elsewhere~\cite{Arleo:2006pc}.

%%%%%%%%%%%%%%%%%%%%%%%%%%%%%%%%%%%%%%%%%%%%%%%%%%%%%%%%%%%%%%%%%%%%%%%%%%%%

\section{Results}
\label{sec:results}

\subsection{Determination of $\sig$ from each experiment}
\label{sec:sigma}

\begin{figure}[htb]  
  \begin{center}
    \includegraphics[height=13.4cm]{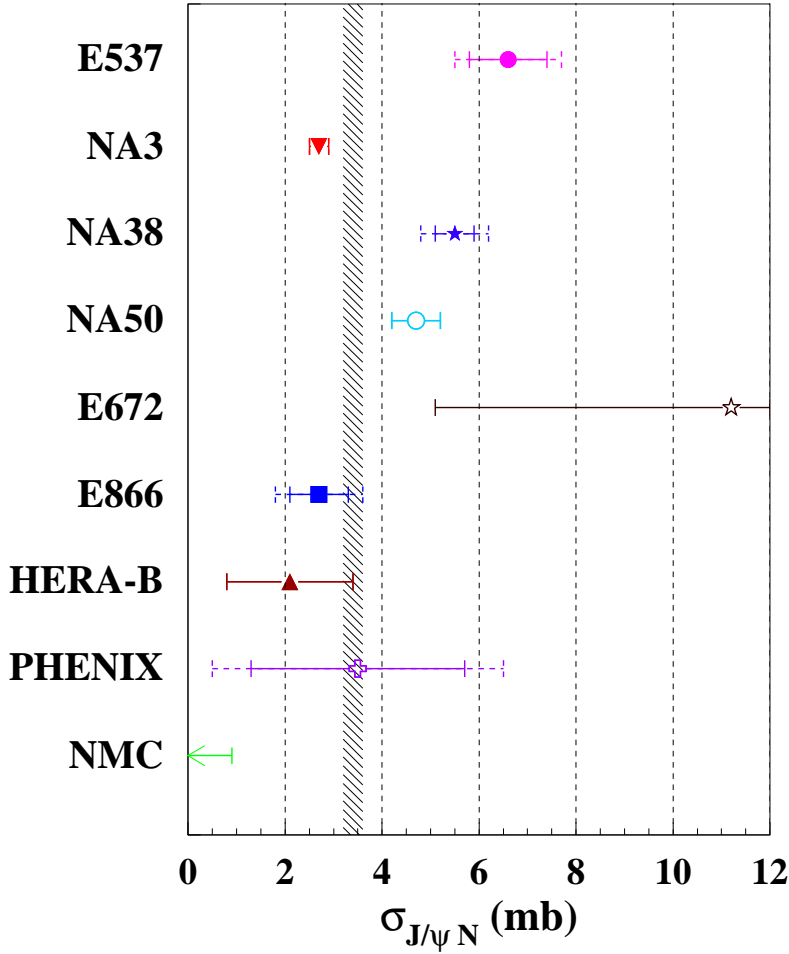}
  \end{center}
\caption{Values of the $\jpsi$--N inelastic cross section extracted from each data set. The solid (dashed) horizontal bars indicate the (rescaled) $1\sigma$ error, $\delta \sig$ ($\delta \sigb$), and the dashed band represents the value obtained from the global fit of all data.}
\label{fig:results}
\end{figure}

The results obtained following the $\chi^2$ analysis described in Sect.~\ref{sec:method} are summarized in Table~\ref{tab:results}. The values of the $\jpsi$--N cross section extracted from the individual fits of each data sample, as well as their corresponding (rescaled) error, $\chi^2$ per number of degrees of freedom ($\chi^2/\ndf$) and S factors, are listed in Table~\ref{tab:results}, using either the proton parton density (left) or the nDS nuclear parton density (right). The results using the proton PDF are also plotted in Fig.~\ref{fig:results}, in which the dashed band indicates the $\jpsi$--N cross section determined from the global fit of all data samples.

\vspace{0.4cm}
\begin{table}[htb]
  \centering
  \begin{tabular}[c]{p{1.8cm}|ccc|ccc}
    \hline
    \hline
     Exp. & $\sig$ (mb) & $\chi^2/\ndf$ & S & $\signDS$ (mb) & $\chi^2_{_{\rm nDS}}/\ndf$ & S \\
     \hline
  & & & & & & \\
 E537      &  6.6 $\pm$  1.1 &  2.0 &  1.4 &  7.0 $\pm$  1.2 &  2.1 &  1.5\\
 NA3       &  2.7 $\pm$  0.2 &  1.5 &  1.2 &  2.9 $\pm$  0.2 &  1.3 &  1.2\\
 NA38      &  5.5 $\pm$  0.7 &  3.4 &  1.8 &  5.4 $\pm$  0.7 &  3.5 &  1.9\\
 NA50      &  4.7 $\pm$  0.5 &  0.4 &  {\rm --} &  4.6 $\pm$  0.5 &  0.4 &  {\rm --} \\
 E672      &  11.2 $\pm$  6.1 &  0.6 &  {\rm --} &  10.3 $\pm$  5.8 &  0.7 &  {\rm --} \\
 E866      &  2.7 $\pm$  0.9 &  2.3 &  1.5 &  2.4 $\pm$  0.7 &  1.6 &  1.3\\
 HERA-B    &  2.1 $\pm$  1.3 &  0.3 &  {\rm --} &  2.0 $\pm$  1.3 &  0.2 &  {\rm --} \\
 PHENIX    &  3.5 $\pm$  3.0 &  1.7 &  1.3 &  3.1 $\pm$  2.6 &  1.4 &  1.2\\
 NMC       &     $\le$  0.9 &  0.7 &  {\rm --}&  $\le$  0.8 &  0.8 &   {\rm --}\\
 & & & & & & \\
 \hline
 Global fit&  3.4 $\pm$  0.2 &  1.5 &  1.2 &  3.5 $\pm$  0.2 &  1.4 &  1.2\\
    \hline
    \hline
\end{tabular}
  \caption{The $\jpsi$--N cross sections, $\chi^2/\ndf$ and S factors extracted from each data sample with (right) and without (left) nDS nuclear corrections. The bottom row corresponds to the global fit of all the data.}
  \label{tab:results}
\end{table}

\begin{figure}[htb]
  \begin{center}
    \includegraphics[height=10.1cm]{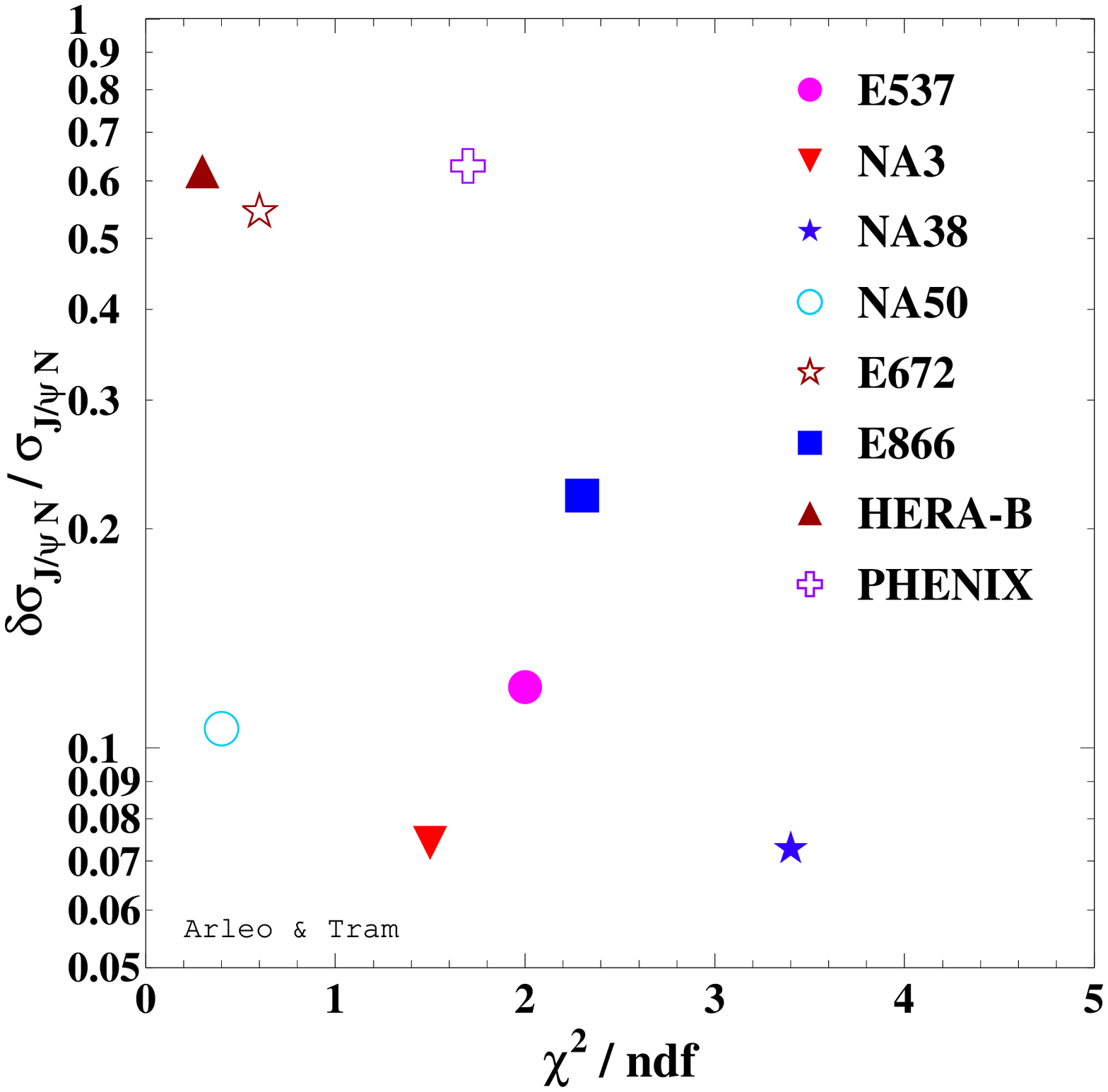}
  \end{center}
\caption{The relative uncertainty of the $\jpsi$--N inelastic cross section, $\delta \sig / \sig$, plotted versus the values of $\chi^2/\ndf$ obtained from each data sample.}
  \label{fig:chi2_dsigma}
\end{figure}

The results obtained without any nuclear modifications in the PDF are fist discussed. From Fig.~\ref{fig:results}, it is clear, first of all, that not all experiments equally constrain the $J/\psi$--N cross section. We may remark in particular that the NA3 and NA50 experiments allow for a precise determination of $\sig$, while the E672, PHENIX and preliminary HERA-B measurements lead to a rather large uncertainty in the parameter estimation. It is also worth while to comment on the agreement between the theoretical calculations and each data set, which is good (with a low $\chi^2/\ndf$) in the E672, NA50, HERA-B and NMC data analysis, but rather poor in the case of the E866 and, especially, NA38 experiments ($\chi^2/\ndf=2.3$ and $\chi^2/\ndf=3.4$, respectively); see Table~\ref{tab:results}. In order to give the reader a feeling on the constraints brought by each experiment and on the agreement between data and theory, the relative uncertainty, $\delta \sig / \sig$, is plotted versus the $\chi^2/\ndf$ value obtained for the various data samples (Fig.~\ref{fig:chi2_dsigma}). We notice, in particular, that the $7\%$ and $11\%$ relative uncertainty in the NA3 and the NA50 data is very good, even though $\chi^2/\ndf$ remains not too large (this is also true, to a lesser extent, with the E537 data). This is in sharp contrast with the E672 and HERA-B experiments, whose agreement with theory is good but which do not constrain $\sig$, and, conversely, with the NA38 experiment for which $\delta \sig / \sig \simeq 7\%$ and $\chi^2/\ndf$ is very large.

\vspace{0.3cm}

We now comment on the values extracted from the various data sets. Surprisingly, one of the most precise determinations of the $\jpsi$--N cross section is due to the NA3 measurements, the ``oldest'' data set analysed here, thanks to the pretty small systematic and statistical error bars on these data points. As can be seen in Fig.~\ref{fig:results}, where all extracted cross sections are plotted, the NA3 fitted value $\sig=2.7\pm 0.2$~mb is perfectly consistent with the estimates from E866 ($\sig=2.7\pm 0.9$~mb), from HERA-B ($\sig=2.1\pm 1.3$~mb), and with the somehow less precise $\sig=3.5\pm 3.0$~mb from the PHENIX experiment. As already pointed out in~\cite{Alessandro:2003pc}, a significantly larger value, $\sig=4.7\pm 0.5$~mb, is determined from the NA50 measurements (this value is consistent with $\sig=4.5\pm 0.5$~mb extracted by NA50 in~\cite{Alessandro:2003pc}). Similarly, a 4-$\delta\sigb$ discrepancy is observed between the NA3 central result and the extracted values from the E537 and NA38 data ($\sig=6.6\pm 1.1$~mb and $\sig=5.5\pm 0.7$~mb); yet the rather poor data--theory agreement in these samples, possibly due to underestimated systematic errors, weakens somehow this observation (Table~\ref{tab:results}). The analysis also shows that the muoproduction data reported by NMC is basically consistent with no $\jpsi$ nuclear absorption, with a not so large $1\sigma$ error, $\delta\sigb=0.9$~mb and a low $\chi^2/\ndf=0.7$. 

\subsection{On correlated systematic errors}
\label{sec:correlated}

The proper statistical treatment of correlated errors in the present $\chi^2$ analysis has been detailed in Sect.~\ref{sec:method}. Neglecting such correlations by adding correlated and uncorrelated errors in quadrature --~yet this is often done in $\chi^2$ fitting procedures~-- indeed leads to strongly biased values for the $\jpsi$--N cross section.

Let us illustrate this with two examples, analysing for instance the E866 and the PHENIX data, which both have normalization (i.e. correlated) errors on the ratio $R^{\rm exp}$ ($\beta=3\%$ and $\beta=12\%$, respectively). The estimate from the E866 experiment goes from $\sig=2.7\pm 0.9$~mb (Table~\ref{tab:results}) to $\sig=3.1\pm 0.3$~mb when correlations are neglected. Similarly, $\sig$ determined from PHENIX is shifted from $\sig=3.5\pm~3.0$~mb to $\sig=2.0\pm~1.3$~mb. The values are thus affected by $15\%$ and $75\%$ --~downwards for E866 and upwards for PHENIX~-- when correlated errors are considered. 

This can be better understood within the ``pull approach'', which is another equivalent way of dealing with correlated systematic errors. When extracting $\sig$ within this alternative approach (with of course identical results), we found that the pulls $\xi$ (that is the amounts, normalized by $\beta$, by which the theory predictions need to be shifted to best accommodate the data) extracted for both experiments indeed are quite large and have opposite signs, $\xi_{_{\rm E866}}=-0.3$ and $\xi_{_{\rm PHENIX}}=0.8$. Perhaps even more importantly, the error on the parameter is smaller by a factor of almost 3 when correlated errors are not properly taken into account.

\subsection{Global fit analysis}
\label{sec:globalfit}

As discussed in Sect.~\ref{sec:sigma}, a significant tension between the different data samples is thus observed. On the one hand, the values extracted from the NA3, E866, HERA-B and PHENIX experiments are somehow consistent with each other (despite some spread), while we remark that the E537, NA38, and NA50 results clearly stand somehow above the bulk of these estimates. The origin of this discrepancy is not clear to us. The vanishing cross section determined from $\jpsi$ muoproduction also stands apart from most of the estimates, even though $\delta\sigb=0.9$~mb. A significantly different nuclear dependence of $\jpsi$ hadroproduction and leptoproduction may actually question the factorization assumed in Sect.~\ref{sec:nucabs}. However, more leptoproduction data would obviously be needed to clarify this issue. 

Despite the above-mentioned tension reported between the individual data sets, the $J/\psi$--N cross section is obtained from a global fit of all data, properly weighted to account for the possible discrepancy between each data sample and the theory (see Sect.~\ref{sec:method}). The global fit leads to a combined cross section of $\sig = 3.4 \pm 0.2$~mb, and naturally a rather large $\chi^2/$ndf $=1.5$. This estimate, as well as its possible energy dependence, will be discussed in Sect.~\ref{sec:discussion}. 

\begin{table}[htb]
  \centering
  \begin{tabular}[c]{cccccc}
 \hline
 \hline
 Excluded expt. &None      &E537      &NA3       &NA38      &NA50      \\
 \hline
 $\sig$ (mb) &$  3.4\pm  0.2$&$  3.3\pm  0.2$&$  4.6\pm  0.3$&$  3.2\pm  0.2$&$
  3.2\pm  0.2$\\
 $\chi^2/\ndf$ &  1.5&  1.4&  1.1&  1.3&  1.5\\
  \hline
  \hline
 Excluded expt. &E672      &E866      &HERA-B    &PHENIX    &NMC       \\
 \hline
 $\sig$ (mb) &$  3.4\pm  0.2$&$  3.4\pm  0.2$&$  3.4\pm  0.2$&$  3.4\pm  0.2$&$
  3.4\pm  0.2$\\
 $\chi^2/\ndf$ &  1.5&  1.5&  1.6&  1.5&  1.5\\
 \hline
 \hline
  \end{tabular}
  \caption{The $\jpsi$--N cross section extracted from the global fit excluding the data from one experiment.}
  \label{tab:excluding}
\end{table}

Given that some experiments give stronger constraints on $\sig$ than others (see above and Fig.~\ref{fig:chi2_dsigma}), it is interesting to repeat the global fit excluding each time one experiment. The values obtained for $\sig$ are quoted in Table~\ref{tab:excluding}. It is not surprising to see that the most dramatic changes on $\sig$ are observed when the NA38 (despite only 3 data points), NA50, and essentially the NA3 experiment (from $\sig=3.4\pm 0.2$~mb to $\sig=4.6\pm 0.3$~mb), relaxing significantly the tension (see Fig.~\ref{fig:results} and the $\chi^2/\ndf$ in Table~\ref{tab:excluding}).

\subsection{Using nuclear parton densities}
\label{sec:nds}

Let us discuss the results of the analysis when the nDS nuclear parton densities are used. We notice in Table~\ref{tab:results} (right) that the inclusion of the nPDF in the theoretical calculation does not significantly affect the results obtained using the proton PDF. Since the shadowing process depletes $\jpsi$ production in nuclei at high energy, the extracted inelastic cross sections tend to be slightly smaller, since part of the $\jpsi$ suppression seen in the data is already accounted for this effect. This is particularly true for the highest-energy E866, HERA-B, and PHENIX experiments. We also note that the agreement between the E866 and the theory is significantly improved, with $\chi^2_{_{\rm nDS}}/\ndf=1.6$ (and a smaller cross section of $\signDS=2.4\pm~0.7$~mb) with respect to $\chi^2/\ndf=2.3$ ($\sig=2.7\pm~0.9$~mb) without any nuclear correction in the PDF. At RHIC, it is also worth while to remark the better agreement when nuclear parton distributions are assumed in the calculation ($\chi^2_{_{\rm nDS}}/\ndf=1.4$) than when they are not ($\chi^2/\ndf=1.7$). At lower energy (larger $x_{_2}$), the effect of nDS antishadowing is rather tiny (see in particular the increase of $\sig$ for the E537 and NA3 data samples); consequently the values of the extracted cross sections are not dramatically affected by these corrections. Note that using nDS PDFs makes $\chi^2/\ndf$ closer to 1 for most experiments. Fitting all data samples results in a cross section $\signDS = 3.5 \pm 0.2$~mb with $\chi^2_{_{\rm nDS}}/\ndf=1.4$, that is perfectly compatible with $\sig = 3.4 \pm 0.2$~mb ($\chi^2/\ndf=1.5$) using the proton densities. 

However, as mentioned in Sect.~\ref{sec:npdf}, the gluon distributions in nuclei are not tightly constrained experimentally (see Fig.~\ref{fig:npdf}). It is therefore the aim of the next section to discuss further the uncertainties of the nPDFs, and especially the consequences on the $\sig$ values extracted from the data.

%%%%%%%%%%%%%%%%%%%%%%%%%%%%%%%%%%%%%%%%%%%%%%%%%%%%%%%%%%%%%%%%%%%%%%%%%%%%

\section{Uncertainties from the nuclear parton densities}
\label{sec:shadowing}

\subsection{Global fit analysis}
\label{sec:globalfitshad}

It has already been stressed that the current DIS and Drell--Yan nuclear measurements do not allow stringent constraints to be put on the amount of gluon (anti)shadowing. Therefore, in order to quantify the sensitivity of our results with respect to the strength of nuclear corrections to the parton densities, the global fit analysis is repeated in this section using other nPDFs parametrizations: nDSg~\cite{deFlorian:2003qf}, EKS~\cite{Eskola:1998iyEskola:1998df}, and HKM~\cite{Hirai:2001npHirai:2004wq}. The results are summarized in Table~\ref{tab:shad}.

\vspace{0.3cm}
\begin{table}[htb]
  \centering
  \begin{tabular}[c]{cccccc}
\hline
\hline
   & Proton& nDS & nDSg & EKS & HKM \\
\hline
$\sigsh$ (mb) & 3.4 $\pm$ 0.2 & 3.5 $\pm$ 0.2 & 3.9 $\pm$ 0.2 & 5.2 $\pm$ 0.2 & 3.5 $\pm$ 0.2 \\
$\chi^2/\ndf$ & 1.5 & 1.4 & 1.5 &  1.6 &  1.9 \\
\hline
\hline
  \end{tabular}
  \caption{The $\jpsi$--N cross section extracted from the data using the proton and various nuclear parton density parametrizations.}
  \label{tab:shad}
\end{table}

The nDSg set is a constrained fit where the gluon shadowing is arbitrarily enhanced, setting $R_g = 0.75$ at $Q^2 = 5$~GeV$^2$ in a Au nucleus. As stressed in~\cite{deFlorian:2003qf}, the agreement between the nDSg parametrization and the DIS and DY data is much worse than the unconstrained fit nDS ($\chi^2_{_{\rm nDSg}}/\ndf=1.4$ versus $\chi^2_{_{\rm nDS}}/\ndf=0.8$), so we use it only to further quantify the systematic uncertainty of the above result, due to the poorly known gluon nPDF. We find a somewhat larger cross section, $\signDSg = 3.9 \pm 0.2$~mb, which is a direct consequence of the gluon antishadowing, with a slightly higher $\chi^2_{_{\rm nDSg}}/\ndf=1.5$. Even more spectacularly, the stronger antishadowing in the EKS parametrization (see Fig.~\ref{fig:npdf}) further enhances the $\jpsi$--N cross section up to $\sigEKS=5.2\pm 0.2$~mb. Finally, the cross section extracted using the HKM parton densities, $\sigHKM=3.5\pm 0.2$~mb, is identical to what is obtained with the nDS parametrization; yet the data--theory agreement is somehow worse ($\chi^2_{_{\rm HKM}}/\ndf=1.9$, as compared with $\chi^2_{_{\rm nDS}}/\ndf=1.4$).

The spread of $\sigsh$ quoted in Table~\ref{tab:shad} directly reflects the present lack of knowledge of the (gluon) nuclear densities. Clearly, a better determination of $\sigsh$ could only be achieved when these are more tightly constrained by experimental data. However, it should be repeated that, among these different parametrizations, the nDS fit offers the best agreement with the nuclear DIS and DY data. In that sense, we believe $\sig^{\rm nDS}=3.5\pm 0.2$~mb to be the most likely estimate. It is also interesting to point out that the $\chi^2$ hierarchy reported here, $\chi^2_{_{\rm nDS}}\lesssim\chi^2_{_{\rm nDSg}}\lesssim\chi^2_{_{\rm EKS}}\lesssim\chi^2_{_{\rm HKM}}$, is the same as what is encountered in the DIS and DY context\footnote{De Florian and Sassot found $\chi^2_{_{\rm nDS}}/\ndf=0.8$, $\chi^2_{_{\rm nDSg}}/\ndf=1.4$, $\chi^2_{_{\rm EKS}}/\ndf=1.6$, and a significantly larger $\chi^2_{_{\rm HKM}}/\ndf\gg 1$ (Ref.~\cite{deFlorian:2003qf}, and private communication).}, although the complexity of the $\jpsi$ production process (as compared with the ``cleaner'' DIS and DY probes) makes it difficult to draw quantitative conclusions on the reliability of the various nuclear distributions. Nevertheless, we shall see in the next section that some experiments actually bring stringent constraints on the nPDFs.

\subsection{Systematics of nuclear parton densities}
\label{sec:moreonnpdf}

For completeness, we give in this section the $\jpsi$--N inelastic cross sections (and $\chi^2/\ndf$) extracted from the individual data samples using the various fits of the nuclear parton densities. In Table~\ref{tab:results_shadowing_all} are listed the estimates using the default nDS fit, as well as the HKM, nDSg, and EKS parametrizations. These last two exhibit a stronger (anti)shadowing than nDS and HKM.

\begin{table}[htb] 
 \centering
{\small
\begin{tabular}[c]{p{1.7cm}|ccc|ccc}
    \hline
    \hline
 & \multicolumn{3}{c}{nDS}& \multicolumn{3}{c}{nDSg} \\
   Exp. & $\signDS$ (mb)& $\chi^2_{_{\rm nDS}}/\ndf$ & S & $\signDSg$ (mb)& $\chi^2_{_{\rm nDSg}}/\ndf$ & S \\
    \hline
 & & & & & & \\
 E537      &  7.0 $\pm$  1.2 &  2.1 &  1.5 &  7.9 $\pm$  1.2 &  2.2 &  1.5 \\
 NA3       &  2.9 $\pm$  0.2 &  1.3 &  1.2 &  3.4 $\pm$  0.2 &  1.4 &  1.2 \\
 NA38      &  5.4 $\pm$  0.7 &  3.5 &  1.9 &  6.2 $\pm$  0.8 &  3.9 &  2.0 \\
 NA50      &  4.6 $\pm$  0.5 &  0.4 &  {\rm --}&  5.4 $\pm$  0.5 &  0.6 &  {\rm --}\\
 E672      &  10.3 $\pm$  5.8 &  0.7 &  {\rm --}&  8.8 $\pm$  5.4 &  0.7 &  {\rm --}\\
 E866      &  2.4 $\pm$  0.7 &  1.6 &  1.3 &  3.6 $\pm$  1.8 &  7.9 &  2.8 \\
 HERA-B    &  2.0 $\pm$  1.3 &  0.2 &  {\rm --}&  2.6 $\pm$  1.4 &  0.3 &  {\rm --}\\
 PHENIX    &  3.1 $\pm$  2.6 &  1.4 &  1.2 &  0.6 $\pm$  1.9 &  0.8 &  {\rm --}\\
 NMC       &  $\le$ 0.8 &  0.8 &  {\rm --} &  $\le$ 0.9 &  0.7 &  {\rm --} \\
 \hline
 Global fit&  3.5 $\pm$  0.2 &  1.4 &  1.2 &  3.9 $\pm$  0.2 &  1.5 &  1.2 \\
   \hline
    \hline
 & \multicolumn{3}{c}{EKS} &        \multicolumn{3}{c}{HKM}\\
   Exp. & $\sigEKS$ (mb)& $\chi^2_{_{\rm EKS}}/\ndf$ & S & $\sigHKM$ (mb)& $\chi^2_{_{\rm HKM}}/\ndf$ & S \\
    \hline
 & & & & & & \\
 E537      &  8.2 $\pm$  1.1 &  1.9 &  1.4 &  7.1 $\pm$  1.1 &  2.0 &  1.4\\
 NA3       &  4.6 $\pm$  0.2 &  1.2 &  1.1 &  3.0 $\pm$  0.2 &  1.5 &  1.2\\
 NA38      &  7.9 $\pm$  0.8 &  3.2 &  1.8 &  5.5 $\pm$  0.7 &  3.4 &  1.8\\
 NA50      &  6.8 $\pm$  0.5 &  0.3 &  {\rm --}&  4.8 $\pm$  0.5 &  0.4 &  {\rm --}\\
 E672      &  11.6 $\pm$  6.3 &  0.6 &  {\rm --}&  10.3 $\pm$  5.8 &  0.6 &  {\rm --}\\
 E866      &  5.3 $\pm$  1.7 &  6.5 &  2.5 &  2.5 $\pm$  0.8 &  2.0 &  1.4\\
 HERA-B    &  4.2 $\pm$  1.5 &  0.9 &  {\rm --}&  2.3 $\pm$  1.3 &  0.2 &  {\rm --}\\
 PHENIX    &  1.3 $\pm$  2.0 &  0.6 &  {\rm --}&  1.5 $\pm$  2.3 &  1.3 &  1.2\\
 NMC       &  $\le$ 1.6 &  0.5 &  {\rm --}&  $\le$  0.9 &  0.7 &  {\rm --}\\
 \hline
 Global fit&  5.2 $\pm$  0.2 &  1.6 &  1.3 &  3.5 $\pm$  0.2 &  1.9 &  1.4\\
    \hline
    \hline
\end{tabular}}
\caption{The $\jpsi$--N inelastic cross section, $\chi^2/\ndf$ and S factors extracted from each data sample using the nDS, nDSg, EKS, and HKM parametrizations for the nuclear PDFs.}
  \label{tab:results_shadowing_all}
\end{table}

As already emphasized, it is in the E866, HERA-B and PHENIX kinematical domain that shadowing corrections prove the strongest. Consequently, the main differences between the various nuclear distributions occur specifically in this region. The results are summarized in Table~\ref{tab:results_shadowing_all}. Overall, the cross sections obtained with HKM are rather similar to those of nDS, while EKS estimates are way different. The nDSg density stands somewhere in between the nDS/HKM and the EKS fitted values. 

It is interesting to note that the cross section extracted at RHIC is basically consistent with~$0$~mb (within 1$\sigma$) using the nDSg, EKS\footnote{Note that the 0--3~mb range used by R.~Vogt for RHIC phenomenology~\cite{Vogt:2004dhVogt:2005ia} is perfectly consistent with our fitted value, $\sigEKS=1.3\pm 2.0$~mb.}, and HKM distributions, while a finite (yet with a large $\delta\sig$) $\signDS=3.1\pm 2.6$~mb, is found when the nDS nPDF is assumed. Let us also mention that the $\chi^2/\ndf$ extracted is much smaller with nDSg and EKS ($\chi^2_{_{\rm nDSg}}/\ndf=0.8$ and $\chi^2_{_{\rm EKS}}/\ndf=0.6$) --~hence for the distributions with the strongest (anti)shadowing~-- than with nDS and HKM ($\chi^2_{_{\rm nDS}}/\ndf=1.4$ and $\chi^2_{_{\rm HKM}}/\ndf=1.3$). 

However, this statement needs to be balanced. Indeed, we can remark that there is a good data--theory agreement (say, with $\chi^2/\ndf\simeq 1$--$2$) for all these nuclear PDFs with each data sample, with the notable exception of the E866 measurements, which strongly disfavour the nDSg ($\chi^2_{_{\rm nDSg}}/\ndf=7.9$) and the EKS ($\chi^2_{_{\rm EKS}}/\ndf=6.5$) distributions and which, in return, definitely support a tiny (anti)shadowing \`a la nDS, with a significantly lower $\chi^2_{_{\rm nDS}}/\ndf=1.6$. Recall that neglecting any nuclear corrections in the PDF yields $\chi^2/\ndf=2.3$ (Table~\ref{tab:results}), that is an intermediate value between nDS on the one hand, and EKS/nDSg on the other hand. Hence, it appears that the E866 $\jpsi$ data give interesting constraints on the evolution of the gluon density ratio, $R_g^{\rm W}(x, Q^2\sim \mpsi^2)$, in the $x=10^{-2}$--$10^{-1}$ range.

\begin{figure}[htb]
  \begin{center}
    \includegraphics[height=10.1cm]{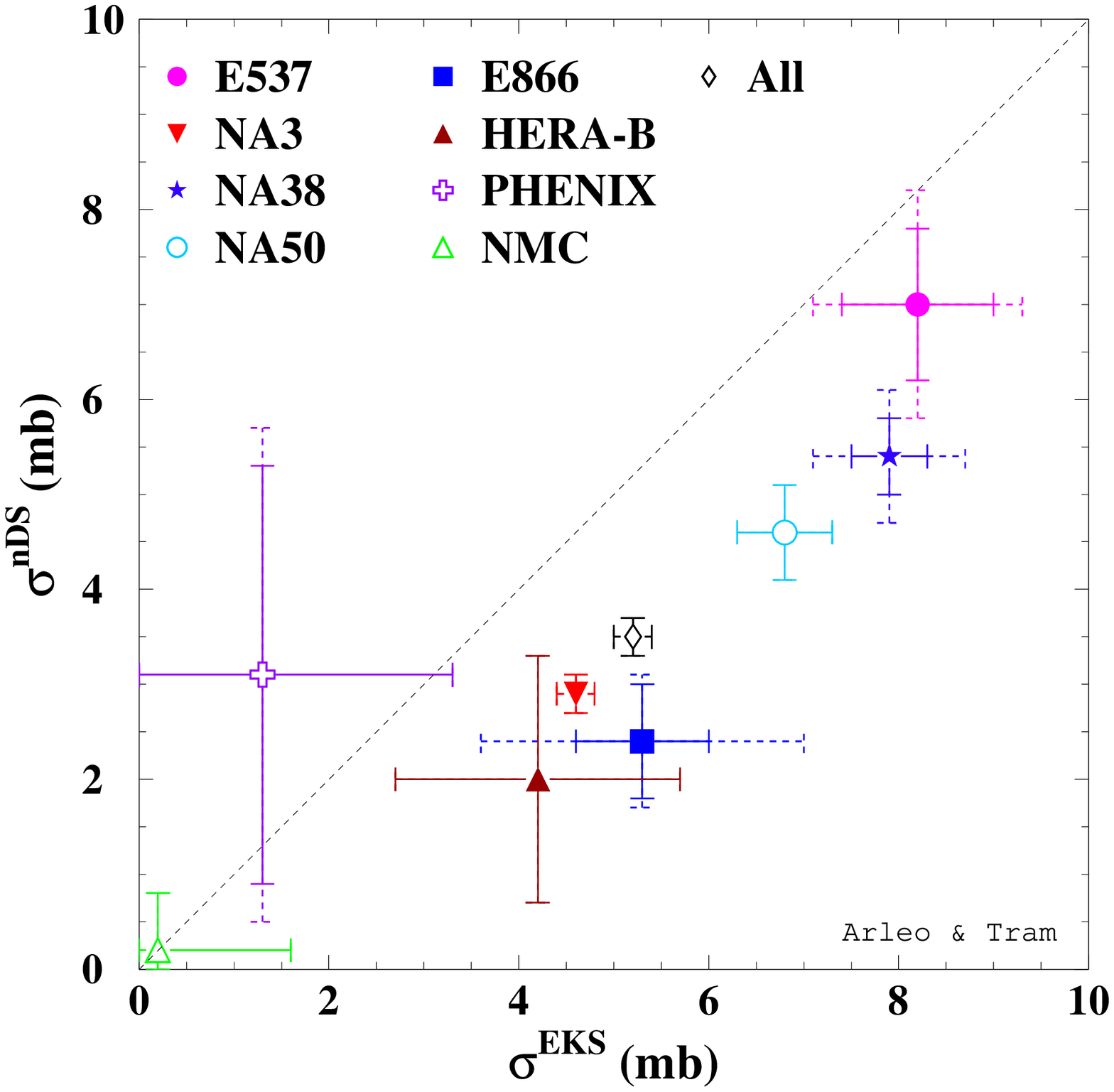}
  \end{center}
\caption{The $\jpsi$--N cross section extracted from each data set and in the global fit, using the nDS ($\signDS$) versus EKS ($\sigEKS$) nuclear parton densities.}
  \label{fig:eksnds}
\end{figure}

In order to stress further the uncertainties on $\sig$ from the unconstrained gluon nPDF, $\signDS$ is plotted as a function of $\sigEKS$ in Fig.~\ref{fig:eksnds}. We remark a systematic offset of roughly $\sigEKS-\signDS\simeq 2$~mb for all but one data sample. As mentioned above, only the PHENIX experiment leads to values almost twice as large with the nDS than with the EKS nPDF. As a consequence, the PHENIX $\sigEKS$ stands significantly below the other lower-energy experiments (with the exception of NMC), while the PHENIX $\signDS$ does not depart much from the NA3, HERA-B and E866 estimates.

%%%%%%%%%%%%%%%%%%%%%%%%%%%%%%%%%%%%%%%%%%%%%%%%%%%%%%%%%%%%%%%%%%%%%%%%%%%%

\section{Discussion}
\label{sec:discussion}

We have presented in the previous section the values extracted for the $\jpsi$--N inelastic cross section from hadroproduction and muoproduction data on nuclear targets, using the simplest geometrical assumptions of the Glauber model. The combined cross section has been found to be $\sig = 3.4 \pm 0.2$~mb, and a slightly larger value $\signDS = 3.5 \pm 0.2$~mb was extracted when the nDS nuclear modifications to the PDFs were taken into account. In this section, we would like first to compare these results with previous phenomenological global analyses in the literature, and then with theoretical expectations. The possible energy dependence of the $\jpsi$--nucleon interaction is finally addressed in Sect.~\ref{sec:energydependence}.

\subsection{Comparing with previous global analyses}
\label{sec:comparison}

In Ref.~\cite{Gerschel:1993uh}, the global analysis of $\jpsi$ suppression in photon--nucleus (E691) and hadron--nucleus (E537, NA3, E672, E772) reactions is carried out by Gerschel and H\"ufner (GH). Excluding the NA3 measurements from the global fit, they found that $\sig=6.2\pm 0.3$~mb. A somewhat larger but compatible result, $\sig=7.3\pm 0.6$~mb, was later determined from the NA38 and E772 measurements of $\jpsi$ production in proton--nucleus collisions by Kharzeev et al. (KLNS) in Ref.~\cite{Kharzeev:1996yx}. These results are significantly larger than our present estimate. Let us try to clarify the origin of this discrepancy. 

The E537, NA3 and E672 data were analysed in both the GH work and the present paper. However, although the values extracted by GH from the NA3 measurements are small (around 3~mb) and compatible with our result, these data were not included in the GH global fit. Moreover, the E691 photoproduction data, which gave $\sig=6.0\pm~1.6$~mb in GH, were not selected in this work (see Sect.~\ref{sec:data}). More importantly, the most accurate estimate of the $\jpsi$--N cross section found in~\cite{Gerschel:1993uh} is due to the E772 measurements ($\sig=6.25\pm 0.5$~mb). However, as already discussed, these data suffer from a bias in the experimental acceptance which systematically increases the nuclear dependence of the $\jpsi$ suppression. The corrected E772 results would then yield a somewhat smaller value for the $\jpsi$--N cross section, which should be compatible with $\sig=2.7\pm 0.9$~mb presently extracted from the more recent E866 experiment in the same kinematic range. Last but not least, many new experimental data (chronologically NMC, NA50, E866, PHENIX, and HERA-B) came out since the analysis of Ref.~\cite{Gerschel:1993uh}. Surprisingly, all these results lead to $\jpsi$--N cross sections smaller, or even much smaller (see Table~\ref{tab:results}), than the estimate quoted in GH. 

Comparing our analysis with that of KLNS, we can remark that the $\sig$ currently extracted from NA38 data ($\sig=5.5\pm 0.7$~mb) is compatible with the value estimated in~\cite{Kharzeev:1996yx}, $\sig=6.2\pm 0.3$~mb. However, the strong disagreement between the data and the theoretical calculations ($\chi^2/\ndf=3.4$) weakens the confidence in this estimate and marginally affects the result from the global fit because of the smaller weight; see Eq.~(\ref{sec:results}).

\subsection{Comparing with phenomenological estimates}
\label{sec:pheno}

Let us now compare our result with phenomenological expectations. The $\jpsi$--N cross section has been computed in a generalized vector dominance model (VDM) in~\cite{Hufner:1997jg}. It is found to be $2.8 \le \sig^{{\rm VDM}} \ \left(\sqrt{s_{_{{\jpsi N}}}} = 10 \ {\rm GeV}\right) \le 4.1 \ {\rm mb}$,
with a smooth energy dependence 
\begin{equation}
  \label{eq:vdm_sdep}
\sig^{{\rm VDM}} \ (\sqrt{s_{_{{\jpsi N}}}}) \ \propto \left( \sqrt{s_{_{{\jpsi N}}}} \right)^\lambda,
\end{equation}
where the exponent of the power-law behaviour $\lambda\simeq 0.4$ is determined from the energy dependence of the $\jpsi$ photoproduction cross section.  Given that the bulk of the data points analysed here lie in a range $\sqrt{s_{_{{\jpsi N}}}}\simeq 10$--20~GeV (see Sect.~\ref{sec:hadro}), the VDM estimate in this energy window then reads
\begin{equation}
  \label{eq:vdm}
2.8 {\rm -} 3.7 \le \sig^{{\rm VDM}} \ \left(\sqrt{s_{_{{\jpsi N}}}} = 10 {\rm -} 20 \ {\rm GeV}\right) \le 4.1 {\rm -} 5.4 \ {\rm mb},
\end{equation}
in perfect agreement with the above extracted value $\sig = 3.4 \pm 0.2$~mb. Note that a similar energy dependence is expected in the pQCD approach by Bhanot and Peskin~\cite{Peskin:1979vaBhanot:1979vb}, where the exponent $\lambda$ is now directly related to the behaviour of the gluon distribution in the nucleon at small $x$. However, it appears that the magnitude of the $\jpsi$--N cross section in this approach is roughly $\sig=1.5$--$2$~mb in the energy domain of interest~\cite{Arleo:2001mp}, that is a factor of 2 smaller than what was found here. Finally, let us mention that the non-perturbative estimate $\sig=3.6$~mb determined in Ref.~\cite{Gerland:1998bz} matches our result well. 

In comparing our result with theoretical expectations, however, we should bear in mind that $\jpsi$ production may also come from the radiative decays of higher states such as $\chi_c$ and $\psi'$ which, because of their larger size, are expected to interact more strongly with the nuclear medium than does the $\jpsi$. In that respect, the measurement of the nuclear dependence of $\chi_c$ production together with a more accurate determination of the $\chi_c$ and $\psi'$ contributions to $\jpsi$ production, possibly performed by the NA60 and HERA-B experiments, is highly desirable (see~\cite{Vogt:2001ky} for a discussion on this issue).

\subsection{Energy dependence}
\label{sec:energydependence}

We have implicitly assumed so far that the $\jpsi$'s are produced instantaneously in the nucleus. This picture is obviously too naive because of the typical proper time $\tau_{_{\jpsi}} \simeq (\mpsip - \mpsi)^{-1}\simeq 0.3$~fm needed to form the $\jpsi$ state~\cite{Brodsky:1988xz,Farrar:1990ei}. At high energy, the Lorentz-contracted nuclear radius, $R/\gamma$, becomes smaller than $\tau_{_{\jpsi}}$; what propagates through the nuclear medium is thus rather a compact $c\bar{c}$ state than a fully formed $\jpsi$ meson. Several models based on colour transparency have been proposed to properly describe the propagation of an expanding $c\bar{c}$ dipole in nuclei~(see e.g.~\cite{Farrar:1990ei,Kopeliovich:1991pu}). In our approach --~which we want to keep as model-independent as possible~-- as well as in Refs.~\cite{Gerschel:1993uh,Kharzeev:1996yx}, such formation time effects are not addressed. Hence, the constant cross section $\sig$ that enters Eq.~(\ref{eq:supp}) has to be seen and understood rather as an {\it effective} absorption parameter (resulting from the average of the $c\bar{c}$ and charmonium interaction with nucleons) than the genuine $\jpsi$--N cross section~\cite{Brodsky:1988xz}, despite the above-mentioned good agreement between our extracted value and the theoretical estimates. 

\vspace{0.3cm}
\begin{figure}[htb]
  \begin{center}
    \includegraphics[height=10.1cm]{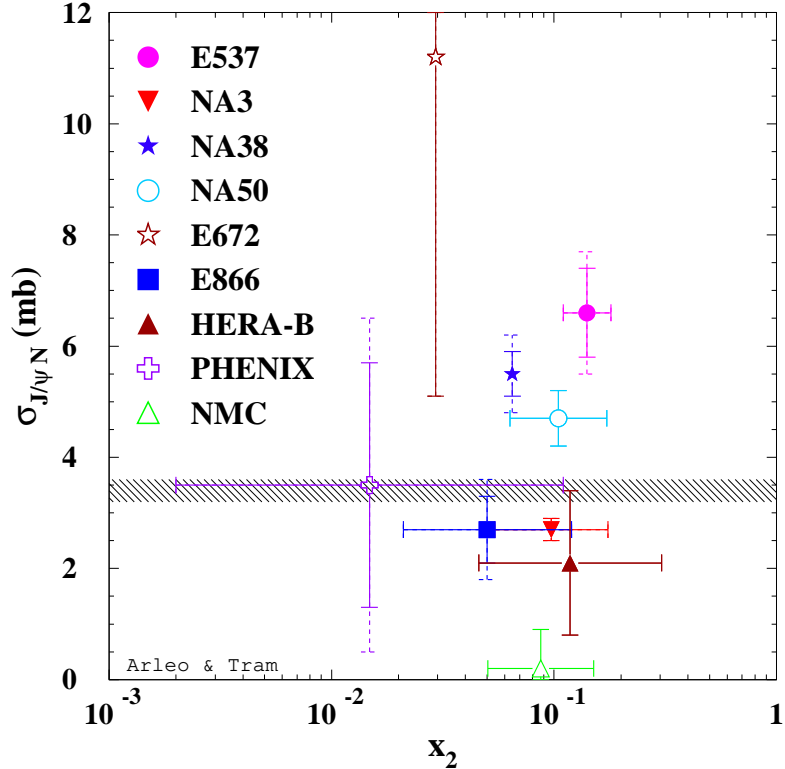}
  \end{center}
\caption{The $\jpsi$--N cross section extracted from each data set as a function of $x_{_2}$. The solid (dashed) vertical bars indicate the (rescaled) 1$\sigma$ error, $\delta \sig$ ($\delta \sigb$), and the dashed band represents the value obtained from the global fit of all data.}
  \label{fig:xsx2}
\end{figure}

Since the $\jpsi$--N energy, Eq.~(\ref{eq:sqrts_psin}) (as well as the boost, $\gamma=\mpsi/ 2\mn x_{_2}$, in the nucleus rest frame), is directly related to the momentum fraction $x_{_2}$, we could --~at least in principle~-- expect the cross section to be a scaling function of $x_{_2}$, $\sig=f(x_{_2})$. In QCD, a compact $c\bar{c}$ dipole does not interact strongly with the nuclear matter because of colour transparency~\cite{Brodsky:1988xz,Farrar:1988me}. Consequently, $f(x_{_2})$ should be a {\it decreasing} function as $x_{_2}$ tends to 0 (i.e. in the high energy limit); yet this trend could somehow be balanced by the energy dependence of Eq.~(\ref{eq:vdm_sdep}), which predicts $\sig \sim \sqrt{s_{_{{\jpsi N}}}}^\lambda \sim x_{_2}^{-\lambda/2}$. 

In order to investigate the possible energy dependence of the $c\bar{c}$--N interaction, we plot in Fig.~\ref{fig:xsx2} the values extracted from the various data samples as a function of $x_{_2}$. The discrepancy between the E537/NA38/NA50 and the NA3/E866/HERA-B experiments already mentioned in Section~\ref{sec:results}  is clearly visible around $x_{_2}\simeq 0.1$. Besides this observation, no real $x_{_2}$-dependence of $\sig$ can be inferred from Fig.~\ref{fig:xsx2}, despite the different dynamics when going from $x_{_2} \simeq 0.2$ down to the $x_{_2} \simeq 10^{-2}$ fraction probed at RHIC\footnote{Recently, an attempt to explain a decrease of the $\jpsi$--N cross section from fixed-target to collider energies has been suggested~\cite{Capella:2006mb}; yet the present analysis indicates that such a decrease may not be statistically significant (cf. Table~\ref{tab:results} and Fig.~\ref{fig:xsx2}).} (this is less true when the EKS nPDF is used, see Sect.~\ref{sec:moreonnpdf}). Needless to mention that having more precise $d$--Au data at RHIC would significantly help to clarify the picture. On a longer term, it would also be interesting to extract similarly $\sig$ from the $\jpsi$ suppression in proton--nucleus collisions at even higher energy (i.e. LHC), so as to eventually signal the onset of colour transparency. Finally, the lack of $x_{_2}$ dependence in the data justifies a posteriori our choice to perform a global fit of the existing measurements with a constant cross section, irrespective of the energy of the reaction. 

%%%%%%%%%%%%%%%%%%%%%%%%%%%%%%%%%%%%%%%%%%%%%%%%%%%%%%%%%%%%%%%%%%%%%%%%%%%%

\section{Summary}
\label{sec:summary}

In summary, all available data on $\jpsi$ hadroproduction and muoproduction on nuclear targets have been analysed within a Glauber model. These measurements allow for the extraction of the $\jpsi$--N inelastic cross section from each experiment. Fitting data altogether leads to $\sig = 3.4 \pm 0.2$~mb ($\chi^2/\ndf=1.5$). A similar analysis, including nuclear corrections in the PDFs, yields comparable results ($\signDS = 3.5 \pm 0.2$~mb, $\chi^2_{_{\rm nDS}}/\ndf=1.4$). We stress however that a much stronger gluon shadowing (such as the one given in nDSg or EKS) leads to a somewhat larger value, even though such a significant antishadowing seems disfavoured by the E866 $\jpsi$ data. 

Remarkably, our result is significantly lower than previous similar estimates~\cite{Gerschel:1993uh,Kharzeev:1996yx}. This is partly the consequence of the new data available (in particular the E866, HERA-B and, to a lesser extent, the PHENIX measurements) which have appeared since the analyses of Refs.~\cite{Gerschel:1993uh,Kharzeev:1996yx}. The cross section found in this paper is rather close to phenomenological expectations for the interaction of a $\jpsi$ state with nucleons. Finally, the present uncertainties of the PHENIX data do not allow a statistically significant energy dependence of $\sig$ to be observed.

%%%%%%%%%%%%%%%%%%%%%%%%%%%%%%%%%%%%%%%%%%%%%%%%%%%%%%%%%%%%%%%%%%%%%%%%%%%%

\section*{Acknowledgements}

We are indebted to Olivier Drapier, Ulrich Husemann and Mike Leitch for useful comments on the NA50, HERA-B and E772/E866 data, and to Daniel de Florian for providing us with the nDS parametrization. Is is also a pleasure to thank David d'Enterria, Rapha\"el Granier de Cassagnac, and St\'ephane Peign\'e for many interesting discussions and suggestions.

%%%%%%%%%%%%%%%%%%%%%%%%%%%%%%%%%%%%%%%%%%%%%%%%%%%%%%%%%%%%%%%%%%%%%%%%%%%%

\providecommand{\href}[2]{#2}\begingroup\raggedright\endgroup

%%%%%%%%%%%%%%%%%%%%%%%%%%%%%%%%%%%%%%%%%%%%%%%%%%%%%%%%%%%%%%%%%%%%%%%%%%%%


\begin{thebibliography}{10}
\bibitem{Matsui:1986dk}
T.~Matsui and H.~Satz, Phys. Lett. {\bf B178} (1986) 416.

\bibitem{Baglin:1990ivBaglin:1991wi}
{\bf NA38} Collaboration, C.~Baglin et al., Phys. Lett. {\bf B220}
  (1989) 471 and {\bf B255} (1991) 459.

\bibitem{Abreu:1997jhAbreu:2000ni}
{\bf NA50} Collaboration, M.~C. Abreu et al., Phys. Lett.
  {\bf B410} (1997) 337 and {\bf B477} (2000)
  28.

\bibitem{Arnaldi:2006ee}
{\bf NA60} Collaboration, R. Arnaldi et al., Nucl. Phys. {\bf A774} (2006) 711.

\bibitem{Adler:2003rcxzAdare:2006ns}
{\bf PHENIX} Collaboration, S.~S. Adler et al., Phys. Rev. {\bf C69} (2004) 014901
  [\href{http://arXiv.org/abs/nucl-ex/0305030}{{\tt nucl-ex/0305030}}];\\
{\bf PHENIX} Collaboration, A. Adare et al., 
  \href{http://arXiv.org/abs/nucl-ex/0611020}{{\tt nucl-ex/0611020}}.

\bibitem{Gunji:2006pc}
{\bf PHENIX} Collaboration, T. Gunji, ``Centrality dependence of $J/\psi$ production in Au -- Au and Cu -- Cu collisions by the PHENIX Experiment at RHIC", talk given at Quark Matter 2006, Shanghai (China), 14-20 Nov 2006.

\bibitem{Peskin:1979vaBhanot:1979vb}
M.~E. Peskin, Nucl. Phys. {\bf B156} (1979) 365;\\
G.~Bhanot and M.~E. Peskin, Nucl. Phys. {\bf B156} (1979) 391.

\bibitem{Kharzeev:1994pz}
D.~Kharzeev and H.~Satz, 
  Phys. Lett. {\bf B334} (1994) 155
  [\href{http://arXiv.org/abs/hep-ph/9405414}{{\tt hep-ph/9405414}}].

\bibitem{Arleo:2001mp}
F.~Arleo, P.~B. Gossiaux, T.~Gousset and J.~Aichelin, Phys. Rev. {\bf D65} (2002)
  014005 [\href{http://arXiv.org/abs/hep-ph/0102095}{{\tt hep-ph/0102095}}].

\bibitem{Hufner:1997jg}
J.~H\"ufner and B.~Z. Kopeliovich, Phys. Lett. {\bf
  B426} (1998) 154 [\href{http://arXiv.org/abs/hep-ph/9712297}{{\tt
  hep-ph/9712297}}].

\bibitem{Katsanevas:1987pt}
{\bf E537} Collaboration, S.~Katsanevas et al., 
   Phys. Rev. Lett. {\bf 60} (1988) 2121.

\bibitem{Badier:1983dg}
{\bf NA3} Collaboration, J.~Badier et al., Z. Phys. {\bf C20} (1983)
  101.

\bibitem{Abreu:1998ee}
{\bf NA38} Collaboration, M.~C. Abreu et al., Phys. Lett. {\bf B444} (1998)
  516.

\bibitem{Alessandro:2003pc}
{\bf NA50} Collaboration, B.~Alessandro et al., Eur.
  Phys. J. {\bf C33} (2004) 31 and \texttt{CERN-PH-EP/2006-018}, to appear in Eur.
  Phys. J. {\bf C}.

\bibitem{Scomparin:2006pc}
{\bf NA60} Collaboration, E.~Scomparin, ``$\jpsi$ production in In--In and $p$--A collisions'', plenary talk given at Quark Matter 2006, Shanghai (China), 14-20 Nov 2006.

\bibitem{Kartik:1990it}
{\bf E672} Collaboration, S.~Kartik et al., Phys. Rev. {\bf D41}
  (1990) 1.

\bibitem{Alde:1990wa}
{\bf E772} Collaboration, D.~M. Alde et al., Phys. Rev. Lett. {\bf 66} (1991)
  133.

\bibitem{Leitch:1999ea}
{\bf E866} Collaboration, M.~J. Leitch et al., Phys. Rev. Lett. {\bf
  84} (2000) 3256 [\href{http://arXiv.org/abs/nucl-ex/9909007}{{\tt
  nucl-ex/9909007}}].

\bibitem{Husemann:2005yq}
{\bf HERA-B} Collaboration, U.~Husemann, 
  DESY-THESIS-2005-005.

\bibitem{Adler:2005ph}
{\bf PHENIX} Collaboration, S.~S. Adler et al., Phys. Rev. Lett. {\bf 96} (2006) 012304
  [\href{http://arXiv.org/abs/nucl-ex/0507032}{{\tt nucl-ex/0507032}}].

\bibitem{Anderson:1976hi}
{\bf SLAC} Collaboration, R.~L. Anderson et al., Phys. Rev. Lett. {\bf 38} (1977)
  263.

\bibitem{Sokoloff:1986bu}
{\bf E691} Collaboration, M.~D. Sokoloff et al., Phys. Rev. Lett. {\bf
  57} (1986) 3003.

\bibitem{Aubert:1984br}
{\bf European Muon Collaboration}, J.~J. Aubert et al., Phys. Lett. {\bf
  B152} (1985) 433.

\bibitem{Amaudruz:1991sr}
{\bf New Muon Collaboration}, P.~Amaudruz et al., Nucl. Phys. {\bf B371} (1992) 553.

\bibitem{Gerschel:1993uh}
C.~Gerschel and J.~H\"ufner, 
  Z. Phys. {\bf C56} (1992) 171.

\bibitem{Kharzeev:1996yx}
D.~Kharzeev, C.~Louren\c{c}o, M.~Nardi and H.~Satz, Z. Phys. {\bf C74} (1997)
  307 [\href{http://arXiv.org/abs/hep-ph/9612217}{{\tt hep-ph/9612217}}].

\bibitem{Barger:1979jsBarger:1980mg}
V.~D. Barger, W.~Y. Keung and R.~J.~N. Phillips, Phys. Lett. {\bf B91} (1980) 253 and Z. Phys. {\bf C6} (1980) 169.

\bibitem{Amundson:1996qr}
J.~F.~Amundson, O.~\'Eboli, E.~M.~Gregores, F.~Halzen, Phys. Lett. {\bf B390} (1997) 323 [\href{http://arXiv.org/abs/hep-ph/9605295}{{\tt hep-ph/9605295}}].

\bibitem{Berger:1980niMartin:1987ww}
E.~L. Berger and D.~L. Jones, Phys. Rev. {\bf D23} (1981) 1521;\\
A.~D. Martin, C.~K. Ng and W.~J. Stirling, 
  Phys. Lett. {\bf B191} (1987) 200.

\bibitem{Pumplin:2002vw}
J.~Pumplin et al., JHEP {\bf 07} (2002) 012
  [\href{http://arXiv.org/abs/hep-ph/0201195}{{\tt hep-ph/0201195}}].

\bibitem{Sutton:1991ay}
P.~J. Sutton, A.~D. Martin, R.~G. Roberts and W.~J. Stirling, Phys. Rev. {\bf D45} (1992) 2349.

\bibitem{Armesto:2006ph}
N.~Armesto, J. Phys. G: Nucl. Part. Phys. {\bf 32} (2006) R367 [\href{http://arXiv.org/abs/hep-ph/0604108}{{\tt hep-ph/0604108}}].

\bibitem{Aubert:1983xm}
{\bf European Muon Collaboration}, J.~J.~Aubert et al., Phys. Lett. {\bf B123} (1983) 275.

\bibitem{Eskola:1998iyEskola:1998df}
K.~J. Eskola, V.~J. Kolhinen and P.~V. Ruuskanen, Nucl. Phys. {\bf B535} (1998) 351
  [\href{http://arXiv.org/abs/hep-ph/9802350}{{\tt hep-ph/9802350}}];\\
K.~J. Eskola, V.~J. Kolhinen and C.~A. Salgado, Eur.
  Phys. J. {\bf C9} (1999) 61
  [\href{http://arXiv.org/abs/hep-ph/9807297}{{\tt hep-ph/9807297}}].

\bibitem{Hirai:2001npHirai:2004wq}
M.~Hirai, S.~Kumano and M.~Miyama, Phys. Rev. {\bf D64} (2001) 034003
  [\href{http://arXiv.org/abs/hep-ph/0103208}{{\tt hep-ph/0103208}}];\\
M.~Hirai, S.~Kumano and T.~H. Nagai, Phys. Rev. {\bf C70} (2004) 044905
  [\href{http://arXiv.org/abs/hep-ph/0404093}{{\tt hep-ph/0404093}}].

\bibitem{deFlorian:2003qf}
D.~de~Florian and R.~Sassot, Phys. Rev. {\bf D69} (2004) 074028
  [\href{http://arXiv.org/abs/hep-ph/0311227}{{\tt hep-ph/0311227}}].

\bibitem{Vogt:2004dhVogt:2005ia}
R.~Vogt, Phys. Rev. {\bf C71} (2005) 054902 [\href{http://arXiv.org/abs/hep-ph/0411378}{{\tt
  hep-ph/0411378}}] and \href{http://arXiv.org/abs/nucl-th/0507027}{{\tt nucl-th/0507027}}.

\bibitem{Capella:1988ha}
A.~Capella, J.~A. Casado, C.~Pajares, A.~V. Ramallo and J.~Tran Thanh~Van, Phys. Lett. {\bf B206} (1988) 354.

\bibitem{DeJager:1987qc}
C.~W. De~Jager, H.~De~Vries and C.~De~Vries, Atom. Data Nucl. Data Tables {\bf 36} (1987) 495.

\bibitem{Pi:1992ug}
H.~Pi, Comput.\ Phys.\ Commun.\ {\bf 71} (1992) 173.

\bibitem{Stump:2001gu}
D. Stump et al., Phys. Rev. {\bf D65} (2002) 014012 [\href{http://arXiv.org/abs/hep-ph/0101051}{{\tt
  hep-ph/0101051}}].

\bibitem{Yao:2006px}
{\bf Particle Data Group}, W.-M. Yao et al., J. Phys. {\bf G33} (2006) 1.

\bibitem{Leitch:2006pc}
M.~Leitch, ``Charmonium production in $p$--A collisions'', Talk given at the International Workshop on the Physics of the Quark--Gluon Plasma, Palaiseau (France), 4--7 September 2001; see \href{http://p25ext.lanl.gov/e866/papers/e866talks/mjlparis01/}{{\tt http://p25ext.lanl.gov/e866/papers/e866talks/mjlparis01/}}.

\bibitem{Vogt:1999dw}
R.~Vogt, Phys. Rev.
  {\bf C61} (2000) 035203 [\href{http://arXiv.org/abs/hep-ph/9907317}{{\tt
  hep-ph/9907317}}].

\bibitem{Arleo:2006pc}
F.~Arleo and D. d'Enterria, under consideration.

\bibitem{Gerland:1998bz}
L.~Gerland, L.~Frankfurt, M.~Strikman, H.~St\"ocker and W.~Greiner, Phys. Rev. Lett. {\bf
  81} (1998) 762 [\href{http://arXiv.org/abs/nucl-th/9803034}{{\tt
  nucl-th/9803034}}].

\bibitem{Vogt:2001ky}
R.~Vogt, Nucl.\ Phys.\ {\bf A700} (2002) 539 [\href{http://arXiv.org/abs/hep-ph/0107045}{{\tt hep-ph/0107045}}].

\bibitem{Brodsky:1988xz}
S.~J. Brodsky and A.~H. Mueller, Phys. Lett. {\bf B206} (1988) 685.

\bibitem{Farrar:1990ei}
G.~R. Farrar, L.~L. Frankfurt, M.~I. Strikman and H.~Liu, Phys. Rev. Lett. {\bf 64}
  (1990) 2996.

\bibitem{Kopeliovich:1991pu}
B.~Z. Kopeliovich and B.~G. Zakharov, Phys. Rev. {\bf D44}
  (1991) 3466.

\bibitem{Farrar:1988me}
G.~R. Farrar, H.~Liu, L.~L. Frankfurt and M.~I. Strikman, Phys. Rev.
  Lett. {\bf 61} (1988) 686.

\bibitem{Capella:2006mb}
A. Capella, E.G. Ferreiro, \href{http://arXiv.org/abs/hep-ph/0610313}{{\tt
  hep-ph/0610313}}.

\end{thebibliography}
\end{document}